\documentclass{article}

\usepackage{emulateapj}
\usepackage{onecolfloat}
\usepackage{graphicx} 
\usepackage{fancyheadings} 
\usepackage{ulem}
\usepackage{rotating}
\usepackage{lscape}

\def\lya{Ly$\alpha$ }
\def\lyb{Ly$\beta$ }

\def\kms{km~s$^{-1}$ }
\def\cm#1{\, {\rm cm^{#1}}}
\def\N#1{{N({\rm #1})}}

\def\rAA{{\rm \AA \, }}
\def\sci#1{{\rm \; \times \; 10^{#1}}}

\def\ohf{{1 \over 2}}
\def\thr{{1 \over 3}}

\def\mkms{{\rm \; km\;s^{-1}}}
\newcommand{\tskip}{\tablevspace{3pt}}

\begin{document}

\twocolumn[%
\submitted{AJ: December 15, 1998}

\title{INVESTIGATING THE METAL LINE SYSTEMS AT z=1.9 TOWARD 
J2233$-$606 IN THE HDF-S\altaffilmark{1}}

\author{ JASON X. PROCHASKA \\
The Observatories of the Carnegie Institute of Washington \\
813 Santa Barbara St. \\
Pasadena, CA 91101 \\
and \\
SCOTT M. BURLES \\
Department of Astronomy and Astrophysics \\
Enrico Fermi Institute \\
University of Chicago \\
5640 S Ellis Ave \\
 Chicago, IL 60637}

\begin{abstract} 

The combination of STIS and
optical spectroscopy with STIS imaging of the field surrounding 
J2233$-$606 afford a unique opportunity to study the physical nature
of the Quasar Absorption Line systems.
We present an analysis of the ionization state, chemical abundances,
and kinematic characteristics of two metal-line systems at
$z=1.92$ and $z=1.94$ toward J2233$-$606.  We focus on these two 
systems because (i) the observations provide full coverage of the
Lyman series, hence an accurate determination of their HI column
densities and (ii) they exhibit many metal-line transitions which
allow for a measurement of their ionization state and chemical abundances.

Line-profile fits of the Lyman series for the two systems
indicate $\log \N{HI} = 17.15 \pm 0.02 \cm{-2}$ evenly distributed
between two components for the $z=1.92$ system
and $\log \N{HI} = 16.33 \pm 0.04 \cm{-2}$ for the $z=1.94$ system.
By comparing observed ionic ratios of C and Si against calculations
performed with the CLOUDY software package, we find the ionization
state is high and well constrained in both systems.
Applying ionization corrections to the measured
ionic column densities, we determine that the systems exhibit significantly
different metallicities: $\approx 1/50$ solar and $< 1/200$ solar
for the two components of $z=1.92$ and $\approx 40 \%$ solar at
$z=1.94$.  The properties of the $z=1.92$ system are consistent with
a low metallicity galaxy (e.g.\ a dwarf galaxy) as well as absorption
by large scale structure of the IGM.
On the other hand, the high metallicity of the $z=1.94$ system suggests
significant star formation and is therefore most likely associated
with a galactic system.  Furthermore, we predict it is the most likely
to be observed with the STIS imaging and follow-up observations.
If a galaxy is identified, our results provide direct measurements
on the properties of the ISM for a $z \approx 2$ galaxy.

\end{abstract}

]

\altaffiltext{1}{Based in part on observations with the NASA/ESA
{\it Hubble Space Telescope} obtained at the Space Telescope Science
Institute, which is operated by AURA, Inc.\ under NASA Contract
NAS 5$-$26555; and observations collected at
European Southern Observatory, La Silla, Chile (ESO Nr. 60.B-0381);
and observations collected at the Anglo-Australian Observatory.}

\pagestyle{fancyplain}
\lhead[\fancyplain{}{\thepage}]{\fancyplain{}{PROCHASKA \& BURLES}}
\rhead[\fancyplain{}{INVESTIGATING THE METAL LINE SYSTEMS AT z=1.9 TOWARD 
J2233$-$606 IN THE HDF-S}]{\fancyplain{}{\thepage}}
\setlength{\headrulewidth=0pt}
\cfoot{}

\section{INTRODUCTION}

Identifying the physical nature of the Quasar Absorption Line (QAL)
systems will have significant impact on our understanding of galaxy
formation, chemical evolution, and large scale structure (LSS).
At high redshift, where QAL systems offer one of the most efficient means of 
probing the early universe, any insight is notable.
To really utilize QAL samples as tools to study cosmology
(LSS, dN/dz) and early galaxy formation,
we need to relate the absorbers to astrophysical environments and objects.
With the advancement of modern telescopes and instrumentation, 
in particular high
resolution echelle spectroscopy, observers are now accurately 
examining the physical characteristics of these systems.
For the damped \lya systems, 
the metallicity (\cite{ptt97}), abundance patterns
(\cite{lu96b,pro99a}) and kinematic characteristics (\cite{pro97b,pro98})
have been well studied and have revealed vital clues to their physical nature.

\begin{table*}
\begin{center}
\caption{\centerline 
{\sc Spectroscopic Data} \smallskip \label{tabdata}}
\begin{tabular}{lccccccc}
\tskip \tableline
\tableline \tskip
Instrument/Telescope & Coverage & Resolution &  Source \nl
\tableline \tskip
E230M-STIS/HST & 2300$-$3100~\AA & $R=30000$ & HST HDF-S STIS Team \nl
G430M-STIS/HST & 2300$-$3100~\AA & 0.56 \aa & HST HDF-S STIS Team \nl
UCLES/AAT & 3530$-$4390~\AA & $R=35000$ & Outram et al.\ (1998) \nl
EMMI/NNT  & 4386$-$8270~\AA & 14 km/s & Savaglio (1998) \nl
\tskip \tableline
\end{tabular}
\end{center}
\end{table*}

Recent observations of the Hubble Deep Field South afford a unique
opportunity to investigate the physical nature of the QAL systems.  
With the combination of STIS spectroscopy and STIS imaging
(\cite{gard99,will99})
of the field toward J2233$-$606  ($z_{em} = 2.238$, $B=17.5$; \cite{byle97})
researchers may directly identify
the absorbers responsible for the QAL systems.
In turn, these observations will further develop the relationship between
QAL systems and galaxies or large scale structure at moderate redshift.
In this paper, we investigate the physical properties of 
the two metal line systems at $z \approx 1.9$ toward J2233$-$606.
We choose these two systems because the STIS spectroscopy coupled
with optical echelle spectroscopy of J2233$-$606 (\cite{out98,svgl98}) 
allows one to well constrain the ionization state 
of these systems.  By applying ionization corrections to the measured 
ionic column densities, we present chemical abundances
for Si, C, N, and Fe.
Finally, we comment on the kinematic characteristics of the 
absorption line systems and
speculate on the systems' association with protogalaxies and large
scale structure at high redshift.

In $\S$~2, we discuss the multiple data sets used in this analysis.
Section 3 details our fits to the Lyman series of the two QAL
systems.  The ionic column densities, ionization state, and 
chemical abundances are examined in $\S$~4.  Finally, we speculate
on the impact of these measurements in $\S$~5.

\section{DATA}

In the following analysis we utilize several data sets obtained with
a number of instruments and telescopes (Table~\ref{tabdata}).  
These include ultra-violet spectra taken
with the STIS E230M echelle and G430M grating on board the Hubble Space
Telescope, and optical echelle spectroscopy acquired with 
the AAT and NTT and kindly provided by 
Outram et al.\ (1998) and Savaglio (1998).
In all, this collection of data 
provides continuous coverage from $2275 - 8300$\rAA
with resolution $R \approx 30000$ (except the region from $3100-3500$\rAA 
corresponding to the G430M grating) and moderate signal to noise (S/N).
With the exception of the NTT spectrum where we have adopted
the continuum provided by Savaglio, we have continuum fit the
spectra with low order Legendre polynomials using a package similar
to the Iraf routine {\it continuum}.

\begin{figure}[hb]
\begin{center}
\includegraphics[height=5.3in, width=3.7in,bb = 55 48 557 744]{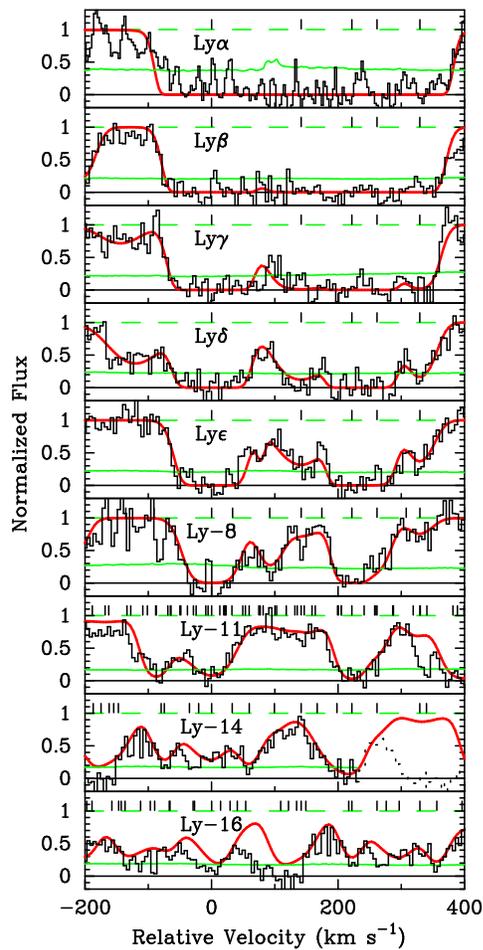}
\caption{A plot of the Lyman series for the Lyman limit system at $z=1.92$
toward J2233$-$606.  Overplotted is the VPFIT solution which includes
several coincident \lya forest clouds.  The dotted regions indicate
data not included in our fit to the Lyman series.}
\label{figLy192}
\end{center}
\end{figure}

\section{HI MEASUREMENTS}

\begin{figure}[ht]
\begin{center}
\includegraphics[height=5.3in, width=3.7in,bb = 55 48 557 744]{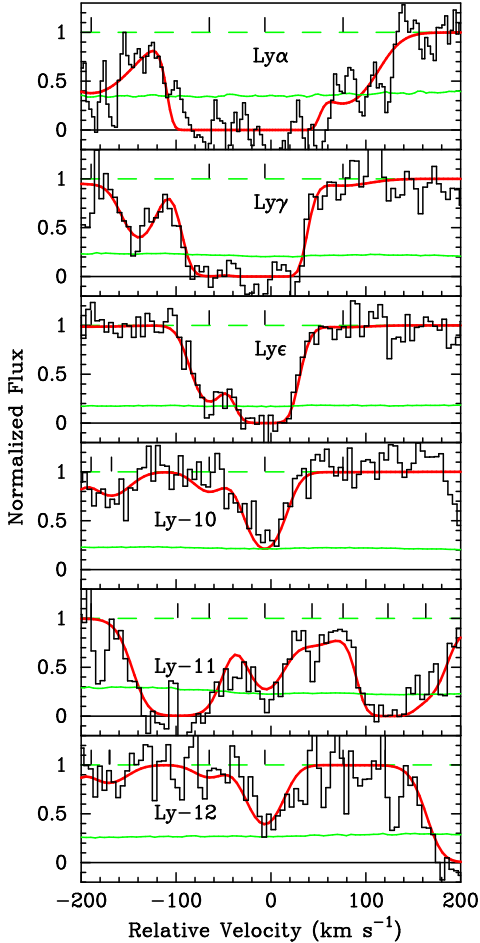}
\caption{A plot of the Lyman series for the QAL system at $z=1.94$
toward J2233$-$606.  Overplotted is the VPFIT solution which includes
several coincident \lya forest clouds.  The dotted regions indicate
data not included in our fit to the Lyman series.}
\label{figLy194}
\end{center}
\end{figure}

A requisite facet in analyzing the ionization state and chemical
abundances of QAL systems is an accurate determination of the HI
column density.  For the QAL systems at $z=1.92$ and $z=1.94$
towards J2233$-$606, the echelle observations performed with STIS
on board the HST are essential as they allow for a fit to the higher
order lines in the Lyman series.  We have measured the HI column
densities by fitting Voigt profiles to the Lyman series
with the VPFIT package kindly provided by R. Carswell and
J. Webb.  Figures~\ref{figLy192} and~\ref{figLy194} and 
Table~\ref{tabLy} present the results of
this analysis.  Note that in performing the fit to the Lyman series, we have
included fits to other \lya forest clouds when necessary.

In the z=1.92 system, most of the H~I is split between two main
components at $z=$1.9256290 and 1.9279202.  
The total column density is $\log \N{HI}  = 17.15 \pm 0.02 \cm{-2}$
with a nearly even split between the main components.  
As suggested by Outram et al.\ (1998), 
the STIS observations clearly demonstrate that the Lyman break evident
at $\approx 2700$~\rAA is due to the Lyman limit (LL) system at $z=1.92$.
For the z=1.94 system, the majority of H~I falls at 
z=1.9425817 and the column density is well constrained by the high-order
unsaturated Lyman lines (Ly-10, 11, 12).  We find 
$\log \N{HI} = 16.33 \pm 0.04 \cm{-2}$.  Therefore this system has a minimal
contribution to the observed Lyman break.

Only weak limits can be placed on D/H in these absorption systems.
In the z=1.94 system, extra H~I falls to the blue of the main H~I 
component (at $-$60 km/s) making future detection of 
the D~I line impossible.
In the z=1.92 system, the \lya line is too wide for a detection or an
upper limit.  The blue wing of \lyb might be used for an upper limit,
but only a weak limit is derived from the 
STIS E230M spectrum: D/H $< 10^{-3}$.

\begin{table} [ht] \footnotesize
\begin{center}
\caption{\label{tabLy}}
{\sc HI FITS FOR THE Z=1.92 AND Z=1.94 QAL SYSTEMS \smallskip}
\begin{tabular}{ccccc}
\tableline
\tableline \tskip
$z_{abs}$ & log $N$  & $\sigma_{log N }$ & $b$ & $\sigma_b$ \\ 
& ($\cm{-2}$) & ($\cm{-2}$) & (\kms) & (\kms) \\
\tableline \tskip
1.2574902 & 12.93 & 0.17  & 8.1  & 4.0  \\
1.2847255 & 13.81 & 0.05  & 48.5 & 6.4  \\
1.3393718 & 13.28 & 0.09  & 42.6 & 10.2 \\
1.3529782 & 13.42 & 0.07  & 21.1 & 3.8  \\
1.4667628 & 13.77 & 0.06  & 24.9 & 2.5  \\
1.9407216 & 13.80 & 0.09  & 49.2 & 6.1  \\
1.9419496 & 15.46 & 0.06  & 20.3 & 1.6  \\
1.9425205 & 16.33 & 0.04  & 22.4 & 1.0  \\
1.9433281 & 13.81 & 0.08  & 37.4 & 8.7  \\
1.9256160 & 16.79 & 0.03  & 33.3 & 1.2  \\
1.9269993 & 15.69 & 0.05  & 44.7 & 4.9  \\
1.9277830 & 16.76 & 0.06  & 21.7 & 1.8  \\
1.9281701 & 16.09 & 0.19  & 23.0 & 4.8  \\
1.9288300 & 15.37 & 0.05  & 25.1 & 1.1  \\
\tskip \tableline
\end{tabular}

\end{center}
\end{table}

\vskip 0.3in 

\section{IONIZATION STATE AND ABUNDANCES}

Since the discovery of QAL metal-line systems,
researchers have investigated the
ionization state of these systems by comparing observations of multiple
ions for a given element against theoretical predictions from photoionization
models (e.g.\ \cite{stdl90,hmmn92}).
While the initial analyses placed meaningful constraints on the ionization
state, they were limited by less accurate column density measurements
derived from lower resolution data.
Recently, Prochaska (1999) has demonstrated that one can
precisely determine the ionization state of a LL system given the 
very accurate column density measurements afforded by modern telescopes.
The combined HST and optical echelle
spectra for J2233$-$606 
provide the opportunity for such an analysis.
In what follows, we investigate the ionization state of 
the two metal-line systems at $z \approx 1.9$ toward J2233$-$606
under the assumption that they are photoionized by
the extragalactic ultra-violet background (EUVB) radiation.
We adopt two input spectra: (i) the Haardt-Madau (1996; HM) EUVB spectrum
which consists entirely of radiation from background quasars
 attenuated by the IGM and
(ii) a model comprised of a synthesis of background galactic and quasar light
(\cite{vrnt99}; VV) also attenuated by the IGM.  The latter model predicts a 
softer spectrum owing to the absorption of radiation with
$h \nu > 4$~Ryd by the ISM of the emitting galaxies.  

Central to our analysis is the CLOUDY software package (V 90.04)
developed and kindly provided by G. Ferland (1995).  We assume
a plane-parallel geometry and approximate 
an isotropic background by placing
a point source at a very large distance.  There is a homologous relation
between the intensity of the EUVB radiation and the volume density of the
QAL system which is typically parameterized via the ionization parameter, $U$.
Here, we define the ionization parameter by relating
the intensity of the EUVB at 1~Ryd (J$_{912}$) with $n_H$:

\begin{equation}
U \equiv {\phi_{912} \over c \, n_H }
= {J_{912} \over 4 \pi h c \, n_H} =
\, (2 \sci{-5}) \; { J_{912} / 10^{-21.5} \over n_H / {\rm cm^{-3}}}
\label{Ueq}
\end{equation}

\noindent  The CLOUDY software package inputs the photoionization spectrum,
$U$, and the observed $\N{HI}$ and then calculates the mean ionization 
fraction for all relevant elements.
This enables one to precisely determine the ionization state of
a QAL system and then apply ionization corrections to the observed ionic column
densities to derive chemical abundances.
We turn now to investigate the QAL systems individually.

\begin{figure}[ht]
\begin{center}
\includegraphics[height=5.3in, width=3.7in,bb = 55 48 557 744]{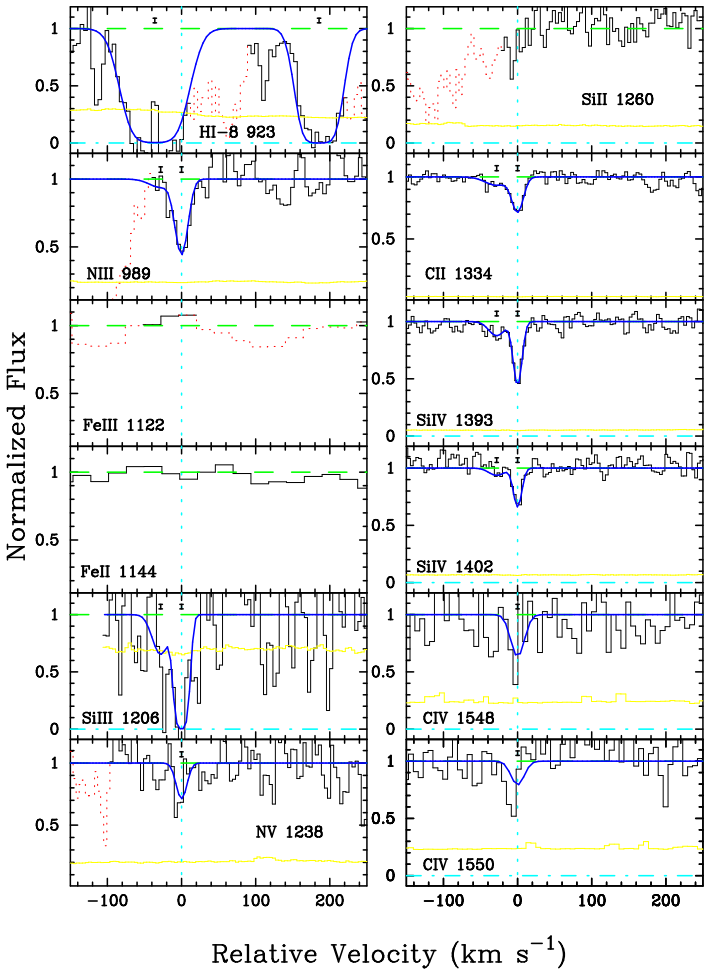}
\caption{Metal-line profiles for the LL system at $z=1.92$.  Overplotted
are our fits to the absorption profiles derived with the VPFIT software
package.  For completeness, we include all metal-line transitions where
one might expect to observe an absorption profile.  We indicate regions
of the profiles contaminated by other absorption lines by dotting the
data.  The error array is the light gray line.}
\label{fig192mtl}
\end{center}
\end{figure}

\vskip 0.2in

\subsection{z=1.92}

This QAL system has a total HI column density $\N{HI} = 10^{17.15} \cm{-2}$
and is responsible for the Lyman limit at $\approx 2700$~\AA.
It exhibits a number of metal-line transitions commonly observed in
LL systems (presented in Figure~\ref{fig192mtl}).  To determine the
ionic column densities from the absorption line profiles, we have
utilized the VPFIT software package in a manner similar to that for
the $\N{HI}$ determinations.
When fitting the profiles we introduced
the minimum number of components necessary to yield a statistically
'good' fit.  
The profiles trace one another very closely 
in velocity space such that we successfully tied the redshifts and 
Doppler parameter of each component for all of the ions.  This point
has significant bearing on our analysis because we believe the
correspondence indicates all of the ions are spatially 
coincident and can be treated together in a single photoionization 
calculation.  Furthermore, it implies there
is little variation in the ionization state along the sightline. 
The VPFIT solutions and $1 \sigma$ errors are presented in 
Table~\ref{tab192vgt}.  With only a few exceptions, the solutions
very closely match the results from Outram et al.\ (1998). 
We have also checked the ionic column density measurements by applying
the Apparent Optical Depth Method (\cite{sav91}) which shows excellent
agreement. For those transitions which exhibit no significant
absorption, we report $3 \sigma$ upper limits to their ionic column
densities.

\begin{table}[hb] \footnotesize
\begin{center}
\caption{\label{tab192vgt}}
{\sc Ionic Column Densities for the $z$=1.92 System \smallskip}
\begin{tabular}{ccccclcc}
\tableline
\tableline \tskip
Comp & $z$ & $\sigma_z$ & b & $\sigma_b$ & Ion & log $N$ &
$\sigma_{\rm{log} {\it N}}$\nl
& & $10^{-6}$ & km/s & km/s & & $\cm{-2}$ & $\cm{-2}$ \nl
\tableline \tskip
 1 & 1.925970 &  3 & 6.13 &  0.41 & Si$^{+3}$ & 12.78 &  0.02 \nl
 & & & 6.13 & & Si$^{++}$  & 13.84 &  0.78 \nl
 & & & 9.37 & & C$^{+}$  & 13.12 &  0.03 \nl
 & & & 9.37 & & C$^{+3}$  & 13.12 &  0.14 \nl
 & & & 8.68 & & N$^{++}$  & 13.79 &  0.12 \nl
 & & & 8.68 & & N$^{+4}$  & 13.06 &  0.15 \nl
 2 & 1.925695 &  14 & 13.89 &  3.00 & Si$^{+3}$ & 12.25 &  0.08 \nl
 & & & 13.89 & & Si$^{++}$  & 12.32 &  0.38 \nl
 & & & 21.24 & & C$^{+}$    & 12.78 &  0.10 \nl
 & & & 19.67 & & N$^{++}$   & 12.93 &  0.62 \nl
\nl
& & & & & Fe$^{+}$ & $< 13.3$ & \nl
& & & & & Fe$^{++}$ & $< 13.5$ & \nl
\tskip \tableline
\end{tabular}
\end{center}
\end{table}

\begin{figure*}[ht]
\begin{center}
\includegraphics[height=5.3in, width=3.7in,angle=-90,bb = 55 48 557 744]{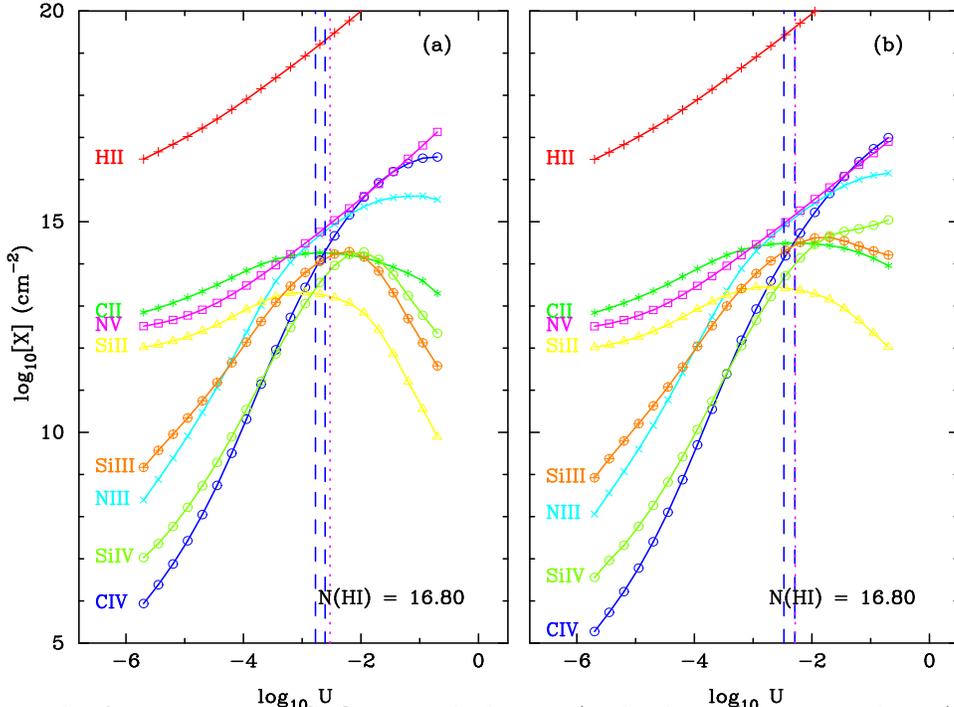}
\caption{The results from a series of CLOUDY calculations (Ferland 1995)
assuming $\log \N{HI} = 16.8$ with (a) a Haardt-Madau ionizing spectrum 
and (b) the Valls-Gabaud/Vernet model.
Plotted are the calculated mean ionic column densities (assuming intrinsic
solar abundances and [Fe/H] = $-0.5$)
versus a modified ionization parameter, $U$, as defined in the text.
The vertical dashed lines indicate the observed C$^+$/C$^{+3}$ ratio
(with errors) and the dotted vertical line is a lower limit to $U$ 
determined from the upper limit to the Si$^+$/Si$^{+3}$ ratio.}
\label{fig192cld}
\end{center}
\end{figure*}

The fact that the HI gas is distributed between two main components
complicates the analysis.  Examining Figure~\ref{fig192mtl}, one notes
that the metal-line profiles lie near the blue HI component ($z=1.9256$), 
albeit not strictly at the center of the profile (compare with the absorption
profiles for the $z=1.94$ system).  Furthermore, we observe no significant 
metal-line profile for the HI component at $z=1.9278$.  
It is very possible that the LL system is actually 
comprised of two spatially separated absorbers with a velocity
separation of $\approx 200 \mkms$ along the sightline.  Because
we cannot constrain the geometry of these absorbers, 
our CLOUDY simulations are an oversimplification.
In the following, we treat the absorbers separately and focus on 
the HI component at $z=1.9256$ with $\N{HI} = 10^{16.79} \cm{-2}$
which corresponds to the metal-line profiles.
Finally, we will use the results of our analysis
to place limits on the metallicity of the other
HI component.  While it is possible that the $z=1.9278$ absorber
may attenuate the EUVB radiation, we find that the ionization corrections
hardly vary for $\log \N{HI} = 16.8 - 17.2 \cm{-2}$ 
and therefore do not consider shielding to be a problem.

\begin{table}[hb] \footnotesize
\begin{center}
\caption{\label{tab192abd}}
{\sc Chemical Abundances for the z=1.92 System \smallskip}
\begin{tabular}{lccccccc}
\tableline
\tableline \tskip
Ion & Haardt/Madau & Valls-Gabaud/Vernet \nl
& [X/H] & [X/H] \nl
\tableline \tskip
C$^+$    & $-1.47 \pm 0.1$  & $-1.70 \pm 0.1$ \nl
C$^{+3}$ & $-1.47 \pm 0.2$  & $-1.70 \pm 0.2$ \nl
N$^{++}$ & $-1.35 \pm 0.2$  & $-1.66 \pm 0.2$ \nl
N$^{+4}$ & $-2.28 \pm 0.3$  & $-2.57 \pm 0.3$ \nl
Si$^+$ & $< -1.54$          & $< -1.68$ 	\nl
Si$^{+3}$& $-1.16 \pm 0.15$ & $-1.43 \pm 0.15$ \nl
Fe$^+$   & $< 1.6$          & $< 1.5$ \nl
Fe$^{++}$& $< -0.1$         & $< -0.2$ \nl
\tskip \tableline
\end{tabular}
\end{center}
\end{table}

Given the ionic column densities and our 
measurement of $\N{HI}$, we can constrain the ionization state
of this system by comparing the relative ionic column densities with
the calculations from the CLOUDY package.  To eliminate the effects of 
the underlying abundance pattern (e.g.\ Type II SN enhancements) and
dust depletion, we will focus on multiple ions from a
single element.  For this LL system, the tightest
constraints are imposed by the C$^+$/C$^{+3}$ ratio.  Figure~\ref{fig192cld}
presents the CLOUDY solutions (calculated ionic column densities)
for a plane-parallel slab with
$\N{HI} = 10^{16.8} \cm{-2}$ photoionized by
(a) an HM EUVB spectrum and (b) the VV spectrum for a range
of ionization parameter assuming [Fe/H]=$-0.5$~dex.  The
results are largely independent of the metallicity assumed in the
CLOUDY calculation; varying [Fe/H] uniformly varies all of the
predicted column densities.
The dashed vertical lines indicate the $U$ values corresponding to
$\log [\N{CII} / \N{CIV}] = 0.16 \pm 0.2$ while the dotted line
denotes a lower limit to $U$ derived from 
$\log [\N{SiII} / \N{SiIV}] < -0.66$.
For the HM spectrum, the C$^+$/C$^{+3}$ and Si$^+$/Si$^{+3}$
ratios are consistent at the $2 \sigma$ level and indicate 
$\log U = -2.69 \pm 0.1$ and $\N{H} = 19.21 \pm 0.20$~dex.
This implies $n_H = 1 \sci{-2} \cm{-3} \, (J_{912}/10^{-21.5})$ which 
agrees with limits to $n_H$ in other 
LL systems (\cite{stdl90,pro99b}).  In terms of the VV EUVB model,
the results are comparable with a slightly higher prediction
for $\N{H}$ and $U$:
$\log U = -2.38 \pm 0.1$, $\N{H} = 19.51 \pm 0.2$~dex, and
$n_H = 5 \sci{-3} \cm{-3} \, (J_{912}/10^{-21.5})$.
Ignoring the effects of clumping within the LL system, 
this implies a characteristic
length scale,  $\ell \sim \N{H} / n_H \sim 1$~kpc.

\begin{figure}[hb]
\begin{center}
\includegraphics[height=5.3in, width=3.7in,bb = 55 48 557 744]{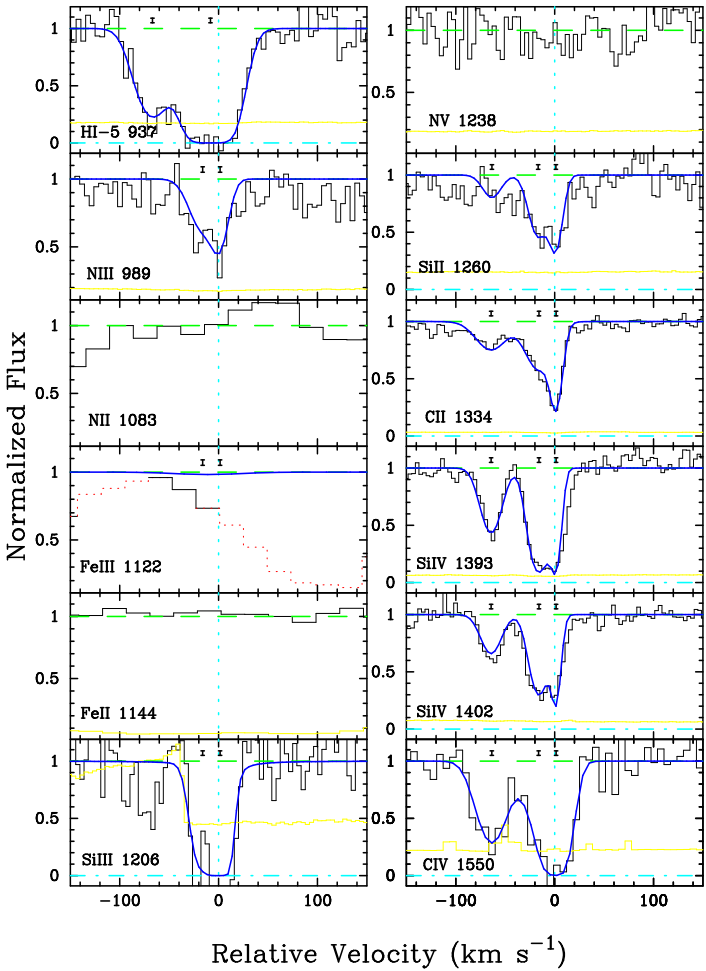}
\caption{Metal-line profiles for the QAL system at $z=1.94$.  Overplotted
are our fits to the absorption profiles derived with the VPFIT software
package.  For completeness, we include all metal-line transitions where
one might expect to observe an absorption profile.  We indicate regions
of the profiles contaminated by other absorption lines by dotting the
data. The error array is the light gray line.}
\label{fig194mtl}
\end{center}
\end{figure}

Having placed relatively tight constraints on the ionization state, we
can calculate the chemical abundances for this system by making
ionization corrections to the observed ionic column
densities.  Table~\ref{tab192abd} lists the abundances relative to solar 
([X/H]~$\equiv$~${\rm \log [\N{X}/\N{H}] - \log [X/H]_\odot}$) derived 
from each
ion for the two EUVB models.  The error bars account for uncertainties
in $\N{HI}$, $U$, and the ionic column density.  
Note the very small error to the [C/H] value derived from $\N{CII}$.
As one observes from Figure~\ref{fig192cld}, the relevant $U$ values
span a very flat section of the CII curve such that
[C/H] is nearly 
independent of $U$.  Therefore, the Carbon abundance
has a very small error to its ionization correction, i.e., the error in
[C/H] is dominated by the uncertainty in the $\N{HI}$ measurement. 
This implies CII measurements can provide reasonably
robust metallicity measurements
for all LL systems with $\N{HI} \approx 10^{17} \cm{-2}$. For this system,
we find [C/H] = $-1.47 \pm 0.1$~dex for the HM spectrum and
[C/H] = $-1.70 \pm 0.1$~dex adopting the VV model.
The values are typical of other estimates for Lyman limit systems at high $z$
(\cite{stdl90,brls98}).  Comparing the Silicon abundance 
([Si/H] = $-1.16 \pm 0.15$ for HM and 
[Si/H] = $-1.43 \pm 0.15$ for the VV model) derived from 
$\N{SiIV}$ against [C/H] we observe good agreement which gives 
further confidence in our modeling as the values are independent
of the C and Si ionic ratios.  The relative abundance of C to Si,
[C/Si] = $-0.3 \pm 0.1$~dex, is consistent with solar abundances at
the $3 \sigma$ level, but favors the ratio of $-0.3$~dex observed in
Population II stars as expected for low metallicity systems.
In all,  we confidently report that the HI component at $z=1.9256$
has a metallicity of $\approx 1/50$ solar abundance.
Without any ionic column density measurements with which to constrain
the ionization state of the HI component at $z=1.9278$, one can only
speculate on its metallicity.  Assuming that the ionization state 
matches that of the blue HI component, we find
[C/H]$< -2.4$~dex from a $3 \sigma$ upper limit to the
CII $\lambda 1334$ column density.

\subsection{z=1.94}

From an examination of a low resolution spectrum of
J2233$-$606, Sealey et al.\ (1998) proposed this QAL system is responsible
for the Lyman limit system observed at $\approx 2700$~\AA.
Figure~\ref{figLy194} clearly demonstrates, however, 
that the Lyman series is not
saturated at Ly-10 and therefore the system is optically  
thin at the Lyman limit.
In Figure~\ref{fig194mtl} we present the velocity profiles 
of the metal-line transitions overplotted
by the VPFIT solution (Table~\ref{tab194vgt}) for those transitions
where an unblended absorption feature is present.  As with
the Lyman limit system at $z=1.92$, the low ionization states
(e.g.\ C$^+$, Si$^+$) very accurately trace the high-ion profiles.
Therefore, we fit all of the profiles by tying
the redshifts and Doppler parameters of the individual components.
Unlike the system at $z=1.92$, the metal-lines also closely trace the HI gas
and we treat all of the gas as a single system.

\begin{figure*}[ht]
\begin{center}
\includegraphics[height=5.3in, width=3.7in,angle=-90,bb = 55 48 557 744]{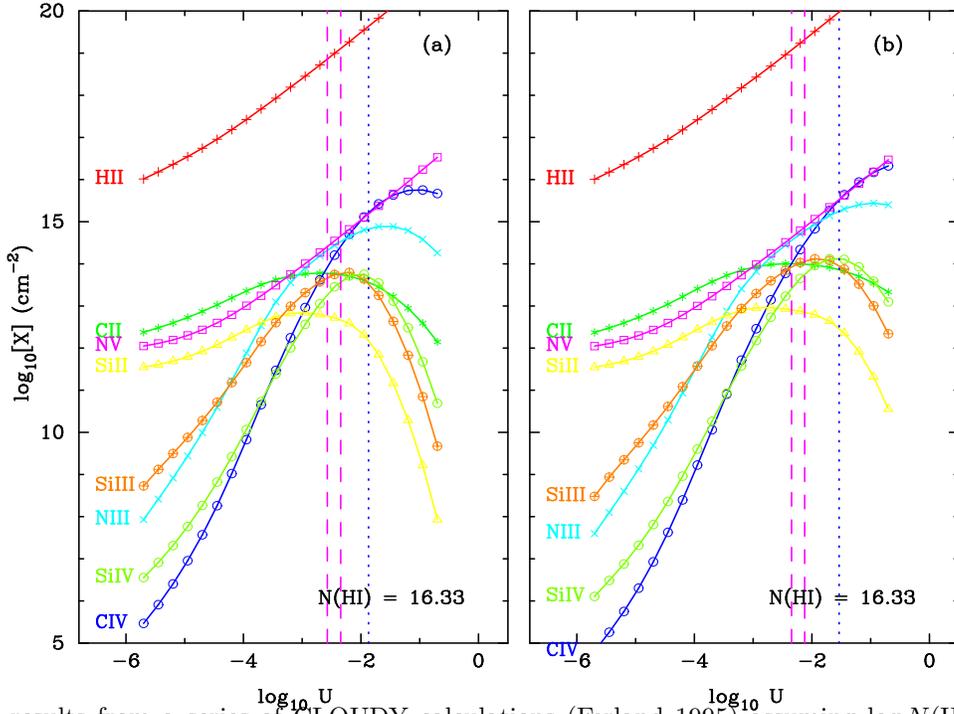}
\caption{The results from a series of CLOUDY calculations (Ferland 1995)
assuming $\log \N{HI} = 16.33$ with (a) a Haardt-Madau ionizing spectrum 
and (b) the Valls-Gabaud/Vernet model.
Plotted are the calculated mean ionic column densities (assuming intrinsic
solar abundances and [Fe/H] = $-0.5$)
versus a modified ionization parameter, $U$, as defined in the text.
The vertical dotted lines indicate the observed Si$^+$/Si$^{+3}$ ratio
(with errors) and the dotted vertical line is a very rough
estimate to $U$ from the C$^+$/C$^{+3}$ ratio.}
\label{fig194cld}
\end{center}
\end{figure*}

\begin{table}[ht] \footnotesize
\begin{center}
\caption{\label{tab194vgt}}
{\sc Ionic Column Densities for the $z$=1.94 System \smallskip} 
\begin{tabular}{ccccclcc}
\tableline
\tableline \tskip
Comp & $z$ & $\sigma_z$ & b & $\sigma_b$ & Ion & log $N$ &
$\sigma_{\rm{log} {\it N}}$\nl
& & $10^{-6}$ & km/s & km/s & & $\cm{-2}$ & $\cm{-2}$ \nl
\tableline \tskip
 1 & 1.942614 &  4 & 6.17 &  0.44 & Si$^{+}$ & 12.69 &  0.12 \nl
 & & & 6.17 & & Si$^{++}$  & 15.03 &  0.81 \nl
 & & & 6.17 & & Si$^{+3}$  & 13.32 &  0.06 \nl
 & & & 9.44 & & C$^{+}$    & 13.63 &  0.04 \nl
 & & & 9.21 & & C$^{+3}$   & 15.35 &  0.56 \nl
 & & & 8.74 & & N$^{++}$   & 13.67 &  0.18 \nl
 2 & 1.942441 &  10 & 11.11 &  0.83 & Si$^{+}$ & 12.71 &  0.10 \nl
 & & & 11.11 & & Si$^{++}$  & 13.06 &  0.43 \nl
 & & & 11.11 & & Si$^{+3}$  & 13.44 &  0.04 \nl
 & & & 17.00 & & C$^{+}$    & 13.56 &  0.05 \nl
 & & & 17.00 & & C$^{+3}$   & 13.80 &  0.27 \nl
 & & & 15.76 & & N$^{++}$   & 13.71 &  0.14 \nl
 3 & 1.941976 &  5 & 11.26 &  0.53 & Si$^{+}$ & 12.15 &  0.20 \nl
 & & & 11.26 & & Si$^{+3}$  & 13.03 &  0.03 \nl
 & & & 17.22 & & C$^{+}$    & 13.26 &  0.03 \nl
 & & & 17.22 & & C$^{+3}$   & 14.04 &  0.14 \nl

\nl
& & & & & N$^{+4}$ & $< 13.3$ & \nl
& & & & & Fe$^{+}$ & $< 13.1$ & \nl
& & & & & Fe$^{++}$ & $< 13.1$ & \nl
\tskip \tableline
\end{tabular}
\end{center}
\end{table}

Figure~\ref{fig194cld} presents the CLOUDY calculations for a
QAL absorption line system with $\N{HI} = 10^{16.33} \cm{-2}$ and
[Fe/H] = $-0.5$~dex adopting (a) a HM EUVB spectrum and
(b) the VV EUVB model.  
Since this system is optically thin at the Lyman limit, the results are
largely insensitive to the $\N{HI}$ value.
Similar to our examination of the $z = 1.92$ system, we overplot the 
allowed values for the ionization parameter given the observed ionic
ratios for Si$^+$/Si$^{+3}$ and C$^+$/C$^{+3}$.  
Because the SiIII and CIV profiles are heavily saturated, the most 
significant constraints for the ionization state are from the 
Si$^+$/Si$^{+3}$ ratio, $\N{SiII}/\N{SiIV} = -0.70 \pm 0.15$~dex.  
For the HM spectrum, we find $\log U = -2.46 \pm 0.1$
which implies $\log \N{H} = 18.98 \pm 0.15 \cm{-2}$ and 
$n_H \approx 6 \sci{-3} \cm{-3} \, (J_{912}/10^{-21.5})$.
Meanwhile, comparing against the CLOUDY solution for the VV EUVB
spectrum, the values are: $\log U = -2.24 \pm 0.1$, 
$\log \N{H} = 19.19 \pm 0.2 \cm{-2}$, and 
$n_H \approx 6 \sci{-3} \cm{-3} \, (J_{912}/10^{-21.5})$. 

Table~\ref{tab194abd} lists the abundances relative to solar for this
system for the two EUVB models.  The variance in the ionization 
corrections for Si$^+$ and C$^+$ are very small ($< 0.1$~dex)
so it is very reassuring that the derived
[C/H] and [Si/H] values are in good agreement:
[C/H] = $-0.27 \pm 0.1$ (HM), [C/H] = $-0.50 \pm 0.1$ (VV) and
[Si/H = $-0.14 \pm 0.1$ (HM), [Si/H] = $-0.31 \pm 0.1$ (VV).
Altogether, the
system has a relatively large metallicity, $\approx \ohf - \thr$ solar 
metallicity.  This value is even larger than
what is typically observed for the damped \lya systems at
$z \approx 2$ (\cite{ptt97,pro99a}).
In $\S$~4 we will discuss the implications of this result.

\begin{table}[ht] \footnotesize
\begin{center}
\caption{\label{tab194abd}}
{\sc Chemical Abundances for the Z=1.94 System \smallskip}
\begin{tabular}{lccccccc}
\tableline
\tableline \tskip
Ion & Haardt/Madau & Valls-Gabaud/Vernet \nl
& [X/H] & [X/H] \nl
\tableline \tskip
C$^+$    & $-0.27 \pm 0.1$  & $-0.50 \pm 0.1$ \nl
C$^{+3}$ & $-0.06 \pm 0.6$  & $-0.15 \pm 0.6$ \nl
N$^{++}$ & $-0.93 \pm 0.2$  & $-1.17 \pm 0.2$ \nl
N$^{+4}$ & $<-3.00 $        & $<-3.15 $ \nl
Si$^+$   & $-0.14 \pm 0.1$  & $-0.31 \pm 0.1$ 	\nl
Si$^{+3}$& $-0.14 \pm 0.1$  & $-0.31 \pm 0.15$ \nl
\tskip \tableline
\end{tabular}
\end{center}
\end{table}

\subsection{HIGHER IONIZATION GAS}

We have also considered constraints on the ionization
state for each system placed by a lower limit
to the N$^{++}$/N$^{+4}$ ratio\footnote{We adopt a lower limit to
this ratio for the $z=1.92$ system
because we expect the NV $\lambda$1238 profile may be
an unidentified, coincident absorption line.}.
The Nitrogen ratios are inconsistent with both
EUVB models for {\it all} values of $U$ in each system.  
We expect this disagreement arises because of a mistreatment
of N$^{+4}$.   N$^{+4}$ is the only ion
we examined with an ionization potential greater than 4 Ryd and is
therefore the most sensitive to three factors: (1) the shape of the
EUVB spectrum at $h \nu > 4$~Ryd, (2) the optical depth of HeII in
the IGM, and (3) the optical depth of HeII within the metal-line system.
Because the number density of quasars peaks near $z = 2$, it is
unlikely that the EUVB spectrum is significantly softer than the VV
model which does account for a contribution from background galaxies.
Regarding the second point, the optical depth of HeII in the IGM is
measured to be quite small at $z = 2.4$ 
($\tau_{HeII} = 1.0 \pm 0.07$; \cite{dvd96}) and is expected to be
significantly lower at $z \approx 2$.  Therefore, we expect the high 
N$^{++}$/N$^{+4}$ ratio implies an oversimplification in modeling
HeII within the CLOUDY framework.  Whereas the CLOUDY package assumes
the metals are uniformly dispersed throughout the absorption-line system, 
it is quite possible the metals are primarily located in the central regions
where they could be shielded from high energy photons by the surrounding
He gas.  For both systems, one predicts $\N{HeII} > 10^{17.8} \cm{-2}$
which imply an optical depth $\tau_{HeII} > 1$ that is
more than enough to account for the $\approx 0.3$~dex 
discrepancy between the observations and CLOUDY predictions.  
Unfortunately, this means one cannot use the NIII or NV 
absorption profiles for constraining the ionization state or abundances 
of this system.  At the same time, however, it is important to stress
that these issues have little effect on the ionization fractions of
C and Si as they are largely insensitive to the flux of photons with
energies $> 4$~Ryd.

\section{SPECULATIONS AND FUTURE OBSERVATIONS}

The similarities and differences observed for the two metal-line
systems toward J2233$-$606 are compelling.  The systems 
differ by only $\approx 0.6$~dex in $\N{HI}$ and have comparable
total Hydrogen column densities.  
The inferred $n_H$ and ionization fractions 
are also in good agreement.
Furthermore, the kinematic characteristics
of the metal-line profiles closely
resemble one another.  The profiles are relatively
narrow ($<100 \mkms$ in velocity space) and exhibit only a modest
asymmetry.  One might expect, therefore, 
that while the $z=1.94$ system may have a slightly lower $\N{H}$ value,
the two systems have a similar physical origin.
There is, however, a significant division between the two systems;
the $z=1.94$ system has a metallicity over 20 times
larger than the $z=1.92$ system.  
We believe this striking comparison 
belies a fundamental difference in the physical nature of the two systems.

To date, several models for the physical nature of
the Lyman limit systems have been proposed.  
Several groups have demonstrated that a significant
fraction of $z \approx 1$ LL systems are associated with the outer
regions (e.g.\ halos)
of $\approx$~L$^*$ galaxies (\cite{berg91,stdl92}).  
This suggests that at $z \approx 2$ one 
expects the Lyman limit systems to be associated with galactic or protogalactic
systems.  On the other hand, numerical hydrodynamic simulations (\cite{ktz96}) 
indicate that LL systems may also arise in the large scale structure
(LSS; e.g.\ filaments, walls) of the Universe.  
The $z=1.92$ system, with its low metallicity,
is a candidate for both of these models.  
If the system is associated with a galaxy, then either very little
star formation (e.g.\ a dwarf galaxy)
has taken place or the sightline intersects
gas in the unenriched outer parts of the galaxy.  
Also, the observation of two significant HI components would indicate
substructure within a single galaxy, or perhaps that the
sightline actually intersects two galaxies. 
The fact that the metal-line profiles do not exactly coincide with either
HI component, however, may be difficult to reconcile in terms of a 
galactic system where one might expect the HI gas to exactly trace 
the CII gas.  At the same time, the observations
can be explained within the framework of a LSS origin.  This scenario, 
similar to the description of the \lya forest within N-body simulations, 
links the Lyman limit system with an unvirialized LSS in the universe, e.g.,
a filament.  The metallicity is sufficiently small that the metals
could have been injected at much earlier time by a galactic system
as is proposed for the enrichment of the \lya forest 
(\cite{tyt95,cwie98,lu98}).
The significant difference in the metallicity for
the two HI components, however, argues for a local production of the
metals (e.g.\ a hydrodynamic shock in the LSS).  While it is not difficult
to envision a LSS scenario which explains the two HI components and the
misalignment of the metal-line profiles, 
more research on this topic
is necessary to reveal the likelihood of these observations. 
Perhaps one will find it is a general prediction from the numerical
simulations that one would not expect the HI gas to directly
align with the metal-line profiles.

In contrast, the high metallicity of
the $z=1.94$ system argues strongly against a LSS origin
as it almost certainly requires the presence of a significant star
forming region.  
Because the system is within 30,000 \kms of the QSO, one must consider
the possibility that it is an associated system.  The fact that
Ly-5 and the CIV profiles are fully saturated indicates no partial
covering and argues strongly against this interpretation.  
Furthermore, no NV absorption is evident as would be expected for
an associated system.
Therefore, we contend the absorber at $z=1.94$ arises within a galactic
system.
In this scenario, the low $\N{H}$ and $\N{HI}$ values indicate
that the sightline has a large impact parameter.  In fact, unless the
system is the result of a very compact star forming region,
we predict one will identify it with the STIS imaging and follow-up
observations. 
If the responsible galaxy is identified, than our analysis affords
a direct measurement of the physical properties of a $z \approx 2$
galaxy.  

By comparing our analysis with an examination of the imaging observations,
one is able to investigate the the ISM in $z \approx 2$ galaxies
(or in the case that no galaxy is identified, perhaps the properties
of LSS).  Unfortunately, one cannot rely on photometric redshifts to
distinguish between $z=1.92$ and $z=1.94$, thus follow-up spectroscopy
of all candidate galaxies will be essential.  
At the time of writing, we are aware of only one 
spectroscopic study (\cite{tresse98}) of the STIS field.  While their
observations identified no galaxies at $z \approx 1.9$, the observations
were limited to I$< 21$.

\acknowledgements

Above all, we wish to acknowledge the many astronomers who have
provided the data for this project.  We wish to thank the HST HDF-S STIS
team (lead by H. Ferguson) for acquiring, reducing, and distributing
the STIS spectroscopy.  We would like to thank P. Outram and
collaborators as well as the AAO for providing their optical spectroscopy.
We also acknowledge S. Savaglio for making public her echelle spectra
of J2233$-$606 obtained with the NTT.
We would also like to acknowledge Bob Carswell and John Webb for
kindly providing the VPFIT package, P. Madau and J. Vernet for providing 
electronic versions of their EUVB spectra and G. Ferland for 
developing and distributing CLOUDY.


\begin{thebibliography}{}

\bibitem[Bergeron \& Boiss$\rm \grave e$ 1991]{berg91}	
Bergeron, J. \& Boiss$\rm \grave e$, P. 1991, \aap, 243, 344

\bibitem[Boyle 1997]{byle97}
Boyle, B.J. 1997, AAO Newsletter, 83, 4

\bibitem[Burles \& Tytler 1998]{brls98}
Burles, S. and Tytler, D. 1998, \apj, 507, 732

\bibitem[Cowie \& Songaila 1998]{cwie98}
Cowie, L. L. \& Songaila, A. 1998, Nature, 394, 44

\bibitem[Davidsen et al.\ 1996]{dvd96}
Davidsen, A. F., Kriss, G. A., \& Zheng, W. 1996, Nature, 380, 47

\bibitem[Ferguson et al.\ 1999]{ferg99}
Ferguson, H.C. et al.\ 1999, \aj, submitted

\bibitem[Ferland 1995]{fer95}
Ferland, G. J. 1995, Hazy, a Brief Introduction to Cloudy, 
Univ. Kentucky, Phys. Dept. Int. Rep

\bibitem[Gardner et al.\ 1999]{gard99}
Gardner, J.P. et al. 1999, \aj, submitted

\bibitem[Haardt \& Madau 1996]{haa96}
Haardt, F. \& Madau, P. 1996, \apj, 461, 20

\bibitem[Hamman \& Ferland 1992]{hmmn92}	
Hamman, F. \& Ferland, G. 1992, \apj, 391, L53

\bibitem[Katz et al.\ 1996]{ktz96}
Katz, N., Weinberg, D.H., Hernquist, L., 
Miralda-Escud$\rm \acute e$, J. 1996, \apj, 456, L57

\bibitem[Lu et al.\ 1996]{lu96b}
Lu, L., Sargent, W.L.W., Barlow, T.A.,
Churchill, C.W., \& Vogt, S. 1996, \apjsupp, 107, 475

\bibitem[Lu et al.\ 1998]{lu98}
Lu, L., Sargent, W. L. W., Barlow, T. A. \& Rauch, M. 1998, submitted
to AJ

\bibitem[Outram et al.\ 1998]{out98}		
Outram, P.J., Boyle, B.J., Carswell, R.F., Hewett, P.C., 
Williams, R.E., \& Norris, R.P. 1998, \mnras, submitted

\bibitem[Pettini et al.\ 1997]{ptt97}
Pettini, M., Smith, L.J., King, D.L., \& Hunstead, R.W. 1997,
\apj, 486, 665

\bibitem[Prochaska 1999]{pro99b}
Prochaska, J.X. 1999, \apj, in press, astro-ph/9811357

\bibitem[Prochaska \& Wolfe 1997b]{pro97b}
Prochaska, J. X. \& Wolfe, A. M. 1997, \apj, 486, 73

\bibitem[Prochaska \& Wolfe 1998]{pro98}
Prochaska, J.X. \& Wolfe, A.M. 1998, \apj, 507, 113

\bibitem[Prochaska \& Wolfe 1999]{pro99a}	
Prochaska, J.X. \& Wolfe, A.M., 1999, \apj, in press, 
astro-ph/9810301

\bibitem[Savage and Sembach 1991]{sav91}
Savage, B. D. and Sembach, K. R. 1991, \apj, 379, 245

\bibitem[Savaglio 1998]{svgl98}			
Savaglio, S. 1998, preprint

\bibitem[Sealey et al.\ 1998]{seal98}		
Sealey, K.M., Drinkwater, M.J., \& Webb, J.K. 1998, \apj, submitted,
astro-ph/9804018

\bibitem[Steidel 1990]{stdl90}
Steidel, C.C. 1990, \apjs, 74, 37

\bibitem[Steidel \& Dickinson 1992]{stdl92}	
Steidel, C.C. \& Dickinson, M. 1992, \apj, 394, 81

\bibitem[Tresse et al.\ 1998]{tresse98}
Tresse, L., Dennefeld, M., Petitjean, P., Cristiani, S., \&
White, S. 1998, \aap, submitted, (astro-ph/9812246)

\bibitem[Tytler et al.\ 1995]{tyt95}
Tytler, D., Fan, X.-M., Burles, S., Cottrell, L., Davis, C., Kirkman,
D., \& Zuo, L. 1995, in QSO Absorption Lines, ed. G. Meylan
(Springer-Verlag), 289

\bibitem[Valls-Gabaud \& Vernet 1999]{vrnt99}
Valls-Gabaud, D. \& Vernet J., \mnras, in preparation

\bibitem[Williams et al.\ 1999]{will99}
Williams, R. et al.\ 1999, \aj, submitted




\end{thebibliography}
\end{document}